# CROR: Coding-Aware Opportunistic Routing in Multi-Channel Cognitive Radio Networks

Xiaoxiong Zhong*, Yang Qin*, Yuanyuan Yang[+], Li Li*
*Key Laboratory of Network Oriented Intelligent Computation, Shenzhen Graduate School,
Harbin Institute of Technology, Shenzhen, 518055, P. R. China
[+]Department of Electrical and Computer Engineering, Stony Brook University, Stony Brook, NY 11794, USA
Email: {xixzhong, yqinsg}@gmail.com, yuanyuan.yang@stonybrook.edu, lili8503@aliyun.com

*Abstract*-Cognitive radio (CR) is a promising technology to improve spectrum utilization. However, spectrum availability is uncertain which mainly depends on primary user's (PU's) behaviors. This makes it more difficult for most existing CR routing protocols to achieve high throughput in multi-channel cognitive radio networks (CRNs). Inter-session network coding and opportunistic routing can leverage the broadcast nature of the wireless channel to improve the performance for CRNs. In this paper we present a coding aware opportunistic routing protocol for multi-channel CRNs, cognitive radio opportunistic routing (CROR) protocol, which jointly considers the probability of successful spectrum utilization, packet loss rate, and coding opportunities. We evaluate and compare the proposed scheme against three other opportunistic routing protocols with multi-channel. It is shown that the CROR, by integrating opportunistic routing with network coding can obtain much better results, with respect to throughput, the probability of PU-SU packet collision and spectrum utilization efficiency.

## I. INTRODUCTION

The cognitive radio principle has introduced the idea to exploit spectrum holes (i.e., bands) which result from the proven underutilization of the electromagnetic spectrum by modern wireless communication and broadcasting technologies [1]. CRNs have emerged as a prominent solution to improve the efficiency of spectrum usage and network capacity. In CRNs, secondary users (SUs) can exploit channels when the primary users (PUs) currently do not occupy the channels. The set of available channels for SUs is instable, varying over time and locations, which mainly depends on the PU's behavior. Thus, it is difficult to create and maintain the multi-hop paths among SUs by choosing both the relay nodes and the available channels to be used on each link of the paths.

Taking the advantage of the broadcast nature and special diversity of the wireless medium, a new routing paradigm, known as opportunistic routing (OR) [2], has been proposed in the ExOR protocol. Instead of first determining the next hop and then sending the packet to it, a node with OR broadcasts the packet so that all neighbors of the node have the chance to hear it and assist in forwarding. OR provides significant throughput gains compared to traditional routing. In CRNs, it is hard to maintain a routing table due to dynamic spectrum access. The pre-determined end-to-end routing does not suit for CRNs. Since opportunistic routing does not need prior setup of the route, it is more suitable for CRNs with dynamic changes of channel availability depending on the PUs behavior.

The effects of opportunistic routing on the performance of CRNs have been investigated in [3-8]. Pan *et al.* [3] proposed a novel cost criterion for OR in CRNs, which leverages the unlicensed CR links to prioritize the candidate nodes and optimally select the forwarder. In this scheme, the network layer selects multiple next-hop SUs and the link layer chooses one of them to be the actual next hop. The candidate next hops are prioritized based on their respective links' packet delivery rate, which in turn is affected by the PU activities. At the same time, Khalife *et al.* [4] introduced a novel probabilistic metric towards selecting the best path to the destination in terms of the spectrum/channel availability capacity. Considering the spectrum available time, Badarneh *et al.* [5] gave a novel routing metric that jointly considers the spectrum availability of idle channels and the required CR transmission times over those channels. This metric aims at maximizing the probability of success (PoS) for a given CR transmission, which consequently improves network throughput. Lin and Chen [6] proposed a spectrum aware opportunistic routing for single-channel CRNs that mainly considers the fading characteristics of highly dynamic wireless channels. The routing metric takes into account transmission, queuing and link-access delay for a given packet size in order to provide guarantee on end-to-end throughput requirement. Taking heterogeneous channel occupancy patterns into account, Liu *et al.* [7] introduced opportunistic routing into the CRN where the statistical channel usage and the physical capacity in the wireless channels are exploited in the routing decision. Liu *et al.* [8] further discussed how to extend OR in multi-channel CRNs based a new routing metric, referred to as Cognitive Transport Throughput (CTT), which could capture the potential relay gain of each relay candidate. The locally calculated CTT values of the links (based on the local channel usage statistics) are the basis for selecting the next hop relay with the highest forwarding gain in the Opportunistic Cognitive Routing (OCR) protocol over multi-hop CRNs.

Network coding (NC) [9] is another technique to increase the throughput of both wired and wireless networks. However, none of the above works systematically combines NC and OR to improve the performance of multi-channel CRNs. Fig.1 shows an example of network coding aware opportunistic routing. In Fig. 1, node 0 wants to send packet $P_1$ to node 2 through the path 0→1→2, and node 7 wants to send packet $P_2$ to node 5 through the path 7 → 6 → 5. These two data transmissions are denoted by green arrow lines. When node 0 sends $P_1$, nodes 1, 3, 4 and 5 will receive/overhear $P_1$ denoted by black arrow lines. Similarly, when node 7 sends $P_2$, nodes 2, 3, 4 and 6 will receive/overhear $P_2$ denoted by black arrow lines. In traditional routing (e.g., AODV, DSR), except for nodes 1 and 6, the remaining nodes will drop the packet received/overheard. While in coding aware OR, it allows nodes 3 and 4 to store packets received/overheard and employ the opportunistic coding approach [10] to code together. Each node in the forwarding set forwards any packets that have still not been received by any higher priority node. If the priority of node 3 is the highest, node 3 XORs $P_1$ and $P_2$, and broadcasts the coded packet $P_1 \oplus P_2$, node 5 and node 2 are able to obtain their needed packets. It requires one additional transmission for transmitting two native packets to node 2 and node 5 without NC. In other words, NC can reduce the number of transmissions in multi-hop wireless networks. Several protocols based on NC scheme have been proposed. COPE [10] is the first practical network coding mechanism for supporting efficient unicast communication in wireless mesh networks (WMNs). In COPE, each node overhears the communications that take place in its neighborhood and records the packets that have been received by the neighbor nodes. In CORE [11], packets of multiple flows are coded together and are transmitted to the candidate next hops. The priority of each candidate is determined in a distributed fashion based on a utility metric. COPE and CORE exploit inter-session network coding to improve the throughput performance of wireless mesh networks. Chachulski *et al.* [12] presented a MAC-independent opportunistic routing & encoding protocol, MORE, which randomly mixes packets before forwarding them based on intra-session network coding. The highlight of MORE is that it does not need any special scheduler to coordinate routers and it runs directly on the top of 802.11 MAC protocol.

All above mentioned coding aware routing protocols are based on single channel networks. It may not be suitable for multi-channel networks, specifically multi-hop CRNs, where the channel availability is uncertain. Note that in our scheme, we mainly consider joint design of inter-session NC and OR in CRNs. Joint design of intra-session NC and multi-channel is also an interesting area. However, it is technically irrelevant to our topic.

The contributions of this paper can be summarized as follows. In this paper, we focus on joint inter-session NC and OR from a successful spectrum utilization perspective in multi-channel CRNs. We first give a method to calculate successful spectrum utilization according to the channel available time and transmission time. And then we introduce a new routing metric, successful delivery ratio (SuDR), which is based on packet loss rate and successful spectrum utilization. Moreover, we propose a novel coding aware OR, cognitive radio opportunistic routing (CROR), based on SuDR metric. Through the evaluation, we validate the effectiveness of the CROR protocol for multi-channel CRNs.

The rest of this paper is organized as follows. Section II introduces the system model. Section III presents metric design concept of SuDR, the method of metric calculation, candidate selection with SuDR and packet encoding algorithm in CROR. Section IV evaluates the performance of CROR by simulations. Finally, section V concludes this paper with a summary of observations.

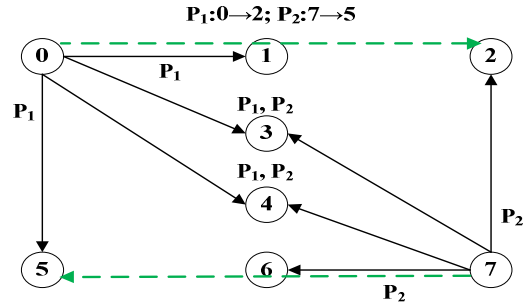

Fig. 1. An example of coding aware OR.

## II. SYSTEM MODEL

In our previous work [13], we assume an interweave model [14], i.e., the nodes in the CRNs can only transmit when the PUs are not active. In this paper, we consider a time slotted multi-hop cognitive radio network (CRN) with $M$ ($M \geq 2$) licensed orthogonal channels. There are $num_s$ SUs and $num_p$ PUs in this CRN. We assume the node (including SU and PU) is equipped with two radios, the one for data transmission and the other one for control signals. Each SU is capable of sensing the locally available channels and has the capability of channel changing at packet level for data transmission.

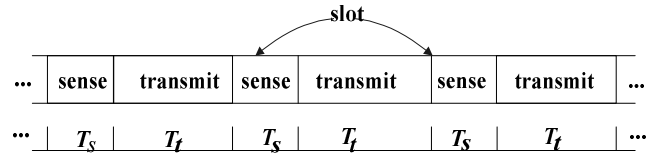

Fig. 2. The time-slotted model for CRNs.

A time-slotted model for SU is assumed, with a fixed slot duration $T$. Each slot consists of a sensing period with duration $T_s$ and a data transmission period with duration $T_t$ ($T_t = T - T_s$), as shown in Fig. 2.

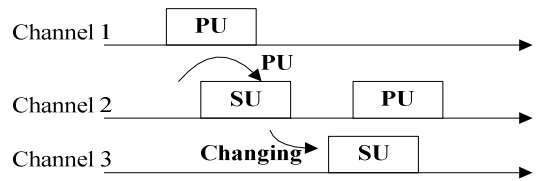

Fig. 3. Spectrum state in spectrum changing for CRNs.

In CRNs, the SUs can opportunistically exploit the channels when the PUs currently do not occupy. When PU appears on a channel, the SU should vacate it for the PU. The SU's communication mainly dependents the PU's activities, which is the biggest difference between CRNs and traditional wireless network, e.g., ad hoc networks and mesh networks. The process of sharing spectrum between PUs and SUs is shown as Fig. 3. The usage pattern of PUs over a given channel $m$ that is available for SU transmissions follows an independent ON/OFF state model [15]. The ON period $T_{on}(m)$ represents the time that the PUs are occupying channel $m$, and the OFF period $T_{off}(m)$ represents the time that the PUs are inactive in channel $m$.

## III. CROR: CODING AWARE OR IN CRNS

The basic idea behind CROR is as follows: first, we propose a new routing metric SuDR, considering available time and packet loss rate of a given channel, for forwarding candidate selection. At each node of a forwarding candidate set (FCS), the packets going to different nexthops are able to decode the resulted packet, and they all share the same channel. The CROR protocol includes following four major parts: 1) new routing metric SuDR; 2) candidate selection and prioritization; 3) coding rules and modeling assumptions; 4) set forwarding timer.

### A. A new routing metric: SuDR

We propose a new metric called successful delivery ratio (SuDR), which captures the number of transmissions needed to deliver a packet for multi-hop CRNs.

MORE and CORE use expected transmission count (ETX) [16] to select forwarding candidate, however, in CRNs, due to PU activities, the routing metric in CRNs must be aware of channel availability at intermediate node. Next, we give the method to calculate SuDR.

The required transmission time for a data packet of size $L$ in channel $m$ between any two neighboring node $i$ and node $j$ (SUs), which can be calculated as follows:

$$T_{tr\_ij}(m) = L / R_{ij}(m) \quad (1)$$

where $R_{ij}(m)$ is the transmission rate in channel $m$.

According to reference [17], we can obtain the channel available time $T_{av\_ij}(m)$, which is exponentially distributed. Thus, the probability $p_{ij}(m)$ that the channel $m$ will be selected successfully for transmission between node $i$ and node $j$ can be represented as

$$p_{ij}(m) = P[T_{av\_ij}(m) \geq T_{tr\_ij}(m)]$$
$$= 1 - F_{T_{av\_ij}(m)}(T_{tr\_ij}(m)) = e^{-\frac{T_{tr\_ij}(m)}{u_{ij}(m)}} \quad (2)$$

where $F_{T_{av\_ij}(m)}(T_{tr\_ij}(m)) = 1 - e^{-\frac{T_{tr\_ij}(m)}{u_{ij}(m)}}$ is the cumulative density function (CDF) of the OFF state of channel $m$ and $u_{ij}(m)$ is channel availability of channel $m$ between node $i$ and node $j$.

Consider the packet loss rate, we calculate the SuDR of a node starting from the destination node, and we set the SuDR value of the destination node as 1. The $SuDR_{ij}$ can be calculated in two adjacent nodes $i$ and $j$, i.e., the number of hops between $i$ and $j$, denoted as $h$, is 1, as follows:

$$SuDR_{ij}(m) = p_{ij}(m) \times (1 - \rho_{ij}(m)) \quad (3)$$

$$SuDR_{ij}[h] = \max_{m \in M}\{SuDR_{ij}(m)\} \quad (h=1) \quad (4)$$

where $M$ is the set of common available channels for node $i$ and node $j$, and $\rho_{ij}(m)$ is the packet loss rate on link between node $i$ and node $j$ on channel $m$.

When the number of hops between $i$ and $j$, $h$, is greater than or equal to 2, the $SuDR_{ij}$ can be recursively expressed as:

$$SuDR_{ij}[h] = 1 - \prod_{r=1}^{N}(1 - SuDR_{ir}SuDR_{rj}[h-1]) \quad (h \geq 2) \quad (5)$$

where $r$ is the node $i$'s neighbor, and $N$ is the number of node $i$'s neighbors. The number of hops between $r$ and $j$ is ($h$-1). Note that the packet forwarding is independent to previous forwarding.

In the following, we give an example to illustrate how the candidate can be selected with ETX and SuDR, respectively. Assume that there are 7 nodes (SUs) and 3 licensed orthogonal channels in CRNs, the network topology is shown as Fig. 4. The available channel set is $M$, $M = \{Ch1,Ch2,Ch3\}$. The available channels of each node are: S= [Ch2, Ch3], 1= [Ch2, Ch3], 2= [Ch1, Ch3], 3= [Ch2, Ch3], 4= [Ch2, Ch3], 5= [Ch1, Ch3], D= [Ch1, Ch2, Ch3]. For simplicity, we assume that the probability $p_{ij}(m)$ is equal between any two neighboring nodes over channel $m$, which is denoted as $p(m)$, also assume that $p(m) = \{p(1), p(2), p(3)\} = \{0.9, 0.92, 0.89\}$. Similarly, the $\rho_{ij}(m)$ can be expressed as $\rho(m)$ over channel $m$, assuming that $\rho(m) = \{\rho(1), \rho(2), \rho(3)\} = \{0.2, 0.3, 0.25\}$.

According to ETX metric and the forwarding scheme, we can obtain the final value of ETX of each node (except for source node S), as shown in Fig. 4 a). Similarly, we can obtain the final value of SuDR of each node (except for source node S), as shown in Fig. 4 b). We observe that the routing metric will affect the path selection. We select the path S-2-5-D with ETX, denoted by red line as shown in Fig. 4 a). However, we select the path S-1-3-D with SuDR, denoted by green line as shown in Fig. 4 b). The SuDR value of a node is its successful delivery ratio from the node to the destination considering the channel availability. The scheme using SuDR metric can achieve better performance than the ones with ETX metric as will be seen in our simulations results later.

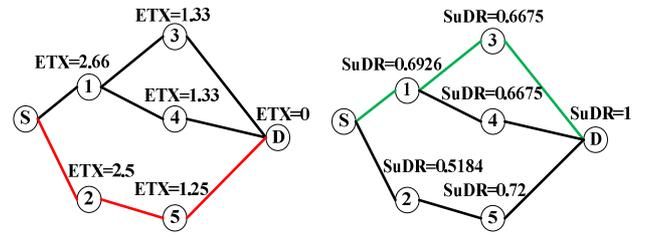

a) Network with ETX metric.  b) Network with SuDR metric.

Fig. 4. Network with different routing metrics.

## B. Candidate selection and prioritization

After detecting an idle channel, the sender needs to select a candidate for data dissemination. Selection of the forwarding nodes must obey the following conditions. The SuDR between the sender and candidates must be larger than the threshold $SuDR_c$ limit to ensure of good quality link.

The **preconditions** that a node can be selected into a FCS are as follows:

(1) *The node and the sender (the node's previous-hop node) share at least one channel; also, the node and its next-hop node share at least one channel.*

(2) *The SuDR value of the node is higher than the threshold.*

(3) *It is a direct neighboring node of the sender.*

(4) *The nodes which are in the FCS must be able to mutually overhear each other with a certain probability.*

The candidate selection algorithm for any node $v$ (except for destination node) is listed in **Algorithm 1**, where $N(v)$ is the set of $v$'s next hop nodes, $Ch(v)$ is the available channel set of node $v$ and $FCS_v$ is the forwarding candidate set of node $v$.

---
**Algorithm 1** Candidate Selection Algorithm

---
1: **function** Selection( $v$, $FCS_v$, $SuDR_c$)
2:   $FCS_v \leftarrow \varnothing$
3:   **for all** node $j \in N(v)$ **do**
4:     Calculate its SuDR according (3), (4), (5)
5:     **if** $(SuDR_j \geq SuDR_c) \&\&(Ch(v)\bigcap Ch(j) \neq \varnothing)$
         $\&\&(Ch(j)\bigcap Ch(N(j)) \neq \varnothing)$ **then**
6:       $FCS_v \leftarrow FCS_v \bigcup j$
7:     **end if**
8:   **end for**
9:   **return** $FCS_v$

---

When a sender is ready to send a packet, it inserts an extra header into this packet, which lists all nodes in FCS. Nodes in the FCS are ranked according to their SuDR. The one with larger SuDR value to the destination has higher priority. Thus, the node will forward packets earlier, and other nodes hearing this forwarding will cancel their timers.

## C. Coding Rules and Modeling Assumptions

Assume packet $P_i$ is at node $v$, $u(P_i)$ is the set of the previous-hop node of packet $P_i$, $n(P_i)$ is the set of the next-hop node of packet $P_i$, $S_v(P_i)$ is the state of packet $P_i$ at node $v$, and $Ch(v)$ is the available channel set of node $v$. If $S_v(P_i) = n$, it means that node $v$ receives the packet $P_i$ in normal fashion. If $S_v(P_i) = d$, it means node $v$ has decoded packet $P_i$. Consider $k$ packets $P_1, P_2, \ldots, P_k$ at node $v$ that have distinct next-hop nodes $n_1, n_2, \ldots, n_k$, respectively. Based on [18], we modify the coding rules, which consider the channel availability and the number of encoding packets. Suppose these are coded together to form the coded packet $P = P_1 \oplus P_2 \oplus \ldots \oplus P_k$ that is broadcast to all the above next-hop nodes. This is a valid network coding if the next-hop node $n_i$ for each packet $P_i$ already has all other packets $P_j$ for $j \neq i$ (so that it can decode $P_i$) - this can occur if

$$n(P_i) = u(P_j) \qquad (6)$$

$$(n(P_i) \in N(u(P_j))) \bigcap S_v(P_j) = n \qquad (7)$$

$$Ch(n(P_i)) \bigcap Ch(n(P_j)) \neq \varnothing \qquad (8)$$

In CROR, all nodes are set in the promiscuous mode, they can overhear packets not addressed to them. Both encoding and decoding are XOR operations. The encoding algorithm is listed in **Algorithm 2**. The parameter average encoding number, $AvgCodingNo_s$, is used to determine how many packets we can encode for practical wireless network coding, XOR, which we can obtain from the reference [19].

---
**Algorithm 2** Encoding Algorithm

---
1: $Capable \leftarrow$ True; $AvgCodingNo \leftarrow 0$; $AvgCodingNo_s \leftarrow 7$
2: $CodedPacket \leftarrow$ front packet of $Q_i$
3: **while** $Capable$ **do**
4:   $Capable \leftarrow$ False
5:   **for all** packet $P_j$ in $Q_i$ **do**
6:     Check the coding conditions according (6), (7), (8)
7:     **If** it satisfies all coding conditions **then**
8:       $Capable \leftarrow$ True
9:       **If** forwarding timer is not expired
           && $P_j$ is not in $CodedPacket$ **then**
10:        $CodedPacket \leftarrow P_j \oplus CodedPacket$
11:        $AvgCodingNo$++
12:      **end if**
13:    **end if**
14:    **If** $AvgCodingNo \geq AvgCodingNo_s$ **then**
15:      $Capable \leftarrow$ False; Break
16:    **end if**
17:  **end for**
18:  Mark latest include native packet as encoded
19: **end while**
20: **return** $CodedPacket$

---

## D. Set forwarding timer

Forwarding timer is the most important aspect of CROR, which is used to avoid redundant transmission, but it also affects the overall throughput of the CROR. After receiving a packet every node sets a timer for each packet. The value of the timer is proportional to the reciprocal of the product of the number of packets that can be coded at the node and the SuDR value. Thus, the node with more coding opportunities and higher SuDR value will forward the packet earlier, and the other nodes hearing this forwarding will cancel their timers.

Let $y$ be the number of packets that can be coded at the node $j$ and $l$ is the order of the node in the FCS. Then the forwarding timer $t$ will be

$$t = \frac{l}{SuDR_j \times y^2} \qquad (9)$$

## IV. PERFORMANCE EVALUATION

In this section, we evaluate the performance of CROR protocol by simulation under different network settings, e.g., PU activity, packet loss rate, and bandwidth difference in spectrum switch state, using NS2 [20] and CRCN model [21]. We set up a CRN with 9 PUs and 32 SUs randomly distributed in a 1500×1500 m² area. As mentioned earlier, the PU activity in each channel is modeled as an exponential ON-OFF process, thus, the channel availability of channel $m$, $\mu_{ij}(m)$ is selected accordingly. The network parameter settings are shown in Table 1. We compare the following three protocols: ExOR [2], MORE [12] and MaxPoS [5], in terms of throughput, the probability of PU-SU packet collision and bandwidth efficiency.

Table 1
Simulation Parameters

| Number of channels | 3 |
|---|---|
| $\mu_{ij}(m)$ (m = 1, 2, 3) | { 0.3; 0.5; 0.7 } |
| Number of PUs per channel | 3 |
| Number of SUs | 32 |
| PU coverage | 550m |
| SU transmission range | 250m |
| Channel data rate | 2Mbps |
| CBR rate | 800Kbps |
| Per channel sensing time | 5ms |
| Channel changing time | 70 $\mu s$ |
| Packet size | 1000 bytes |
| E[$T_{off}$] | [100ms,700ms] |
| AvgCodingNo$_s$ | 7 |
| SuDR threshold | 0.55 |

### A. Impact of PU activity on throughput

In this section, we study the impact of the PU activity on network throughput. When the duration of PU occupation is fixed (5s), as shown in Fig. 5, we observe that with the increase of PU arrival rate, the throughput of CROR, ExOR, MORE and MaxPoS schemes will decrease. On PUs arrival, the affected SUs immediately interrupt transmission and then find an available channel to continue to transmit data packets if any or wait for an opportunity. When the PU arrival rate is larger, it has much more interruption loss for SUs, and then it takes much less time for SUs transmission. The proposed protocol, CROR and MaxPoS outperform other two schemes. This is because that they consider the appropriate channel selection and the probability of successful transmission in SU's communication. However, CROR is better than MaxPoS, this is because the coding opportunities are considered in CROR, which can reduce the number of transmissions, alleviates congestion, and consequently yields higher throughput. In addition, MORE has a better performance than ExOR, this is due to the fact that the source node transmits random linear combination of packets in MORE, which can exploit spatial reuse.

### B. Impact of PU activity on probability of PU-SU packet collision

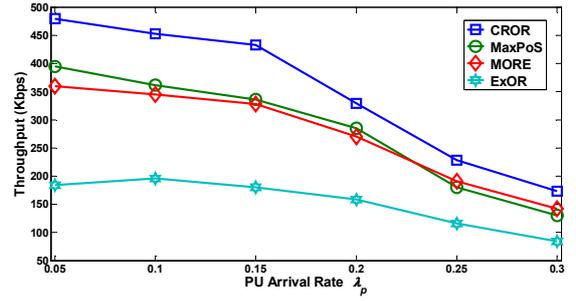

Fig. 5. Throughput vs. PU arrival rate $\lambda_p$.

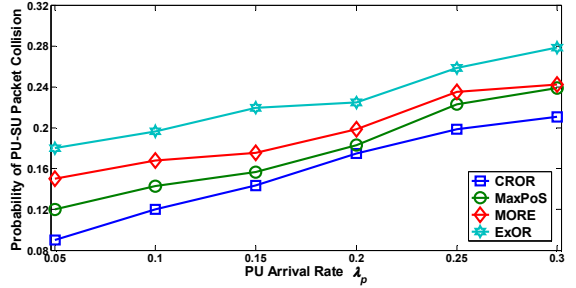

Fig. 6. Probability of PU-SU packet collision vs. PU arrival rate $\lambda_p$.

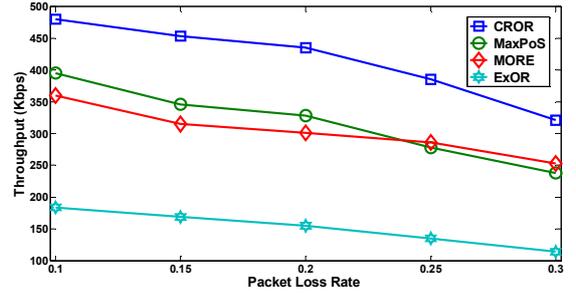

Fig. 7. Throughput vs. packet loss rate.

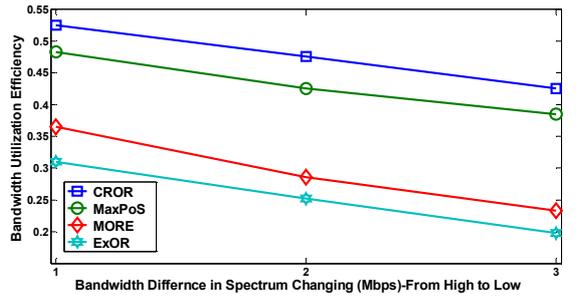

Fig. 8. Bandwidth efficiency vs. difference in spectrum changing.

CROR scheme reduces the probability of PU-SU packet collision. Fig. 6 shows that the interference to SUs increases with the PU arrival rate. The duration of PU occupation is fixed (5s). A routing scheme may choose a route that minimizes the probability of PU-SU packet collisions. CROR and MaxPoS outperform other two schemes which consider the available time and transmission time of the channel for SU's communication. Also, we can see that exploit network coding in routing design can reduce the probability of PU-SU packet collisions (e.g., CROR, and MORE).

## C. Impact of packet loss rate

Next, we evaluate the impact of packet loss rate on throughput. In this test, we vary the packet loss rate between any two nodes which are neighbors per channel. Fig.7 shows that as the packet loss rate increases, the throughput decreases in all routing protocols. This is because the larger packet loss rate, the more packets should be retransmitted. However, the CROR still outperforms other three schemes with higher throughput, which simultaneously exploits channel availability and network coding in routing design. In addition, we can observe that when packet loss rate is larger, the opportunistic routing protocols (CROR, MORE) based network coding have a better performance than that of the protocols without network coding, which means that network coding still brings coding gain in larger packet loss rate with less number of transmissions.

## D. Impact of spectrum changing on bandwidth utilization efficiency

The impact of spectrum changing on bandwidth utilization efficiency is depicted in Fig. 8. On PUs arrival, the affected SUs should switch channel for continuously transmitting data packets. In this process, the bandwidth may change, which mainly affect the transmission time and packet loss rate. We investigate the bandwidth efficiency of the four schemes in this scenario: three channels, Ch1, Ch2, Ch3, having varying raw channel bandwidth: 1Mbps, 2Mbps and 4Mbps. The switch sequences are 2Mbps→1Mbps, 4Mbps→2Mbps, and 4Mbps → 1Mbps. The bandwidth differences are 1Mbps, 2Mbps, and 3Mbps. The number of PU appearance is 1 and the PU on-time is 1s. As seen in Fig. 8, it shows that as the bandwidth difference grows, the bandwidth utilization efficiency reduces. However, CROR has higher bandwidth efficiency than other three schemes, implying that our scheme is effective in fully utilizing the spectrum resource, which considers the channel available time, transmission time and coding opportunities in routing design.

## V. CONCLUSION

In this paper, we propose a novel coding-aware opportunistic routing, CROR, for multi-channel CRNs. In this scheme, we exploit a new routing metric SuDR, which takes channel available time, transmission time and the coding opportunities into consideration. The scheme aims to discover the maximal probability of successful transmission (includes packet loss rate and channel availability) and coding opportunities. Simulation results demonstrate that the proposed routing CROR achieves high performance in increasing the throughput and bandwidth utilization efficiency and reducing the probability of PU-SU packet collision compared with ExOR, MORE and MasPoS. Our future work will focus on design of coding and interference aware opportunistic routing for multi-channel CRNs.

ACKNOWLEDGMENT

This work was supported by the Science and Technology Fundament Research Fund of Shenzhen under grant JC200903120189A,JC201005260183A,ZYA201106070013A.

We would like to acknowledge the reviewers whose comments and suggestions significantly improved this paper.